\documentclass[12pt,a4paper]{article}
\usepackage{amsmath,amssymb,graphicx,epsfig,cite,color}
\usepackage[left=2.7cm,right=2.7cm,top=2.5cm,bottom=2.5cm]{geometry}
\usepackage{subfigure}
\usepackage{xcolor}

\usepackage{graphicx}
\usepackage{lscape}
\usepackage{appendix}
\usepackage{multirow}
\usepackage[colorlinks=true,linkcolor=red]{hyperref}%

\begin{document}

\def\ds{\displaystyle}
\def\beq{\begin{equation}}
\def\eeq{\end{equation}}
\def\bea{\begin{eqnarray}}
\def\eea{\end{eqnarray}}
\def\beeq{\begin{eqnarray}}
\def\eeeq{\end{eqnarray}}

\def\rar{\rightarrow} 
\def\nnb{\nonumber}

\def\ds{\displaystyle}
\def\beq{\begin{equation}}
\def\eeq{\end{equation}}
\def\bea{\begin{eqnarray}}
\def\eea{\end{eqnarray}}
\def\beeq{\begin{eqnarray}}
\def\eeeq{\end{eqnarray}}
\def\ve{\vert}
\def\vel{\left|}
\def\ver{\right|}
\def\nnb{\nonumber}
\def\ga{\left(}
\def\dr{\right)}
\def\aga{\left\{}
\def\adr{\right\}}
\def\lla{\left<}
\def\rra{\right>}
\def\rar{\rightarrow}
\def\lrar{\leftrightarrow}
\def\nnb{\nonumber}
\def\la{\langle}
\def\ra{\rangle}
\def\ba{\begin{array}}
\def\ea{\end{array}}
\def\tr{\mbox{Tr}}
\def\ssp{{\Sigma^{*+}}}
\def\sso{{\Sigma^{*0}}}
\def\ssm{{\Sigma^{*-}}}
\def\xis0{{\Xi^{*0}}}
\def\xism{{\Xi^{*-}}}
\def\qs{\la \bar s s \ra}
\def\qu{\la \bar u u \ra}
\def\qd{\la \bar d d \ra}
\def\qq{\la \bar q q \ra}
\def\gGgG{\la g^2 G^2 \ra}
\def\q{\gamma_5 \not\!q}
\def\x{\gamma_5 \not\!x}
\def\g5{\gamma_5}
\def\sb{S_Q^{cf}}
\def\sd{S_d^{be}}
\def\su{S_u^{ad}}
\def\sbp{{S}_Q^{'cf}}
\def\sdp{{S}_d^{'be}}
\def\sup{{S}_u^{'ad}}
\def\ssp{{S}_s^{'??}}

\def\sig{\sigma_{\mu \nu} \gamma_5 p^\mu q^\nu}
\def\fo{f_0(\frac{s_0}{M^2})}
\def\ffi{f_1(\frac{s_0}{M^2})}
\def\fii{f_2(\frac{s_0}{M^2})}
\def\O{{\cal O}}
\def\sl{{\Sigma^0 \Lambda}}
\def\es{\!\!\! &=& \!\!\!}
\def\ap{\!\!\! &\approx& \!\!\!}
\def\ar{&+& \!\!\!}
\def\ek{&-& \!\!\!}
\def\kek{\!\!\!&-& \!\!\!}
\def\cp{&\times& \!\!\!}
\def\se{\!\!\! &\simeq& \!\!\!}
\def\eqv{&\equiv& \!\!\!}
\def\kpm{&\pm& \!\!\!}
\def\kmp{&\mp& \!\!\!}
\def\mcdot{\!\cdot\!}
\def\erar{&\rightarrow&}



\renewcommand{\textfraction}{0.2}    
\renewcommand{\topfraction}{0.8}   

\renewcommand{\bottomfraction}{0.4}   
\renewcommand{\floatpagefraction}{0.8}
\newcommand\mysection{\setcounter{equation}{0}\section}

\def\baeq{\begin{appeq}}     \def\eaeq{\end{appeq}}  
\def\baeeq{\begin{appeeq}}   \def\eaeeq{\end{appeeq}}
\newenvironment{appeq}{\beq}{\eeq}   
\newenvironment{appeeq}{\beeq}{\eeeq}
\def\bAPP#1#2{
 \markright{APPENDIX #1}
 \addcontentsline{toc}{section}{Appendix #1: #2}
 \medskip
 \medskip
 \begin{center}      {\bf\LARGE Appendix #1 :}{\quad\Large\bf #2}
\end{center}
 \renewcommand{\thesection}{#1.\arabic{section}}
\setcounter{equation}{0}
        \renewcommand{\thehran}{#1.\arabic{hran}}
\renewenvironment{appeq}
  {  \renewcommand{\theequation}{#1.\arabic{equation}}
     \beq
  }{\eeq}
\renewenvironment{appeeq}
  {  \renewcommand{\theequation}{#1.\arabic{equation}}
     \beeq
  }{\eeeq}
\nopagebreak \noindent}

\def\eAPP{\renewcommand{\thehran}{\thesection.\arabic{hran}}}

\renewcommand{\theequation}{\arabic{equation}}
\newcounter{hran}
\renewcommand{\thehran}{\thesection.\arabic{hran}}

\def\bmini{\setcounter{hran}{\value{equation}}
\refstepcounter{hran}\setcounter{equation}{0}
\renewcommand{\theequation}{\thehran\alph{equation}}\begin{eqnarray}}
\def\bminiG#1{\setcounter{hran}{\value{equation}}
\refstepcounter{hran}\setcounter{equation}{-1}
\renewcommand{\theequation}{\thehran\alph{equation}}
\refstepcounter{equation}\label{#1}\begin{eqnarray}}


\newskip\humongous \humongous=0pt plus 1000pt minus 1000pt
\def\caja{\mathsurround=0pt}
 

\title{
         {\Large
                 {\bf
Strong Coupling Constants of the Doubly Heavy $ \Xi_{QQ} $ Baryons with $ \pi $ Meson  
                 }
         }
      }

\author{\vspace{1cm}\\
{\small
A. R. Olamaei$^{1,4}$,  
K. Azizi$^{2,3,4}$, S.~Rostami$^{5}$} \\
{\small$^1$  Department of Physics, Jahrom University, Jahrom, P.~ O.~ Box 74137-66171, Iran}\\
{\small $^2$ Department of Physics, University of Tehran, North Karegar Ave. Tehran 14395-547, Iran}\\
{\small $^3$ Department of Physics, Do\v{g}u\c{s} University, Acibadem-Kadik\"{o}y, 34722
Istanbul, Turkey}\\
{\small $^4$ School of Particles and Accelerators, Institute for Research in Fundamental 
Sciences (IPM),}\\
{\small  P. O. Box 19395-5531, Tehran, Iran}\\
{\small$^5$  Department of Physics, Shahid Rajaee Teacher Training University, Lavizan, Tehran 16788, Iran}} 

\date{}

\begin{titlepage}
\maketitle
\thispagestyle{empty}

\begin{abstract}
The doubly charmed $\Xi_{cc}^{++} (ccu)$ state is the only listed baryon in PDG, which was discovered in the experiment. The LHCb collaboration gets closer to discovering the second doubly charmed baryon $\Xi_{cc}^{+} (ccd)$, hence the investigation of the doubly charmed/bottom baryons from many aspects is of great importance that may help us not only get valuable knowledge on the nature of the newly discovered states, but also in the search for other members of the doubly heavy baryons predicted by the quark model. In this context, we investigate the strong coupling constants among the $\Xi_{cc}^{+(+)}$ baryons and $\pi^{0(\pm)}$ mesons by means of light cone QCD sum rule. Using the general forms of the interpolating currents of the $\Xi_{cc}^{+(+)}$ baryons and the distribution amplitudes (DAs) of the $\pi$ meson, we extract the values of the coupling constants $g_{\Xi_{cc} \Xi_{cc} \pi}$. We extend our analyses to calculate the strong coupling constants among the partner baryons with $\pi$ mesons, as well, and extract the values of the strong couplings $g_{\Xi_{bb} \Xi_{bb} \pi}$ and $g_{\Xi_{bc} \Xi_{bc} \pi}$.  The results of this study may help experimental groups in the analyses of the data related to the strong coupling constants among the hadronic multiplets.

\end{abstract}

\end{titlepage}
\section{INTRODUCTION}

The search for doubly heavy baryons and determination of their properties constitute one of the main directions of the research in the experimental and theoretical high energy physics. There is only one doubly charmed baryon, $\Xi_{cc}^{++}$, listed in the PDG. The searches for other members of the doubly heavy baryons in the experiments, as the natural outcomes of the quark model, are in progress. Theoretical investigations on properties of the doubly heavy baryons, are necessary as their results can help us better understand their structure and the dynamics of the QCD as the theory of the strong interaction.

The search for doubly heavy baryons is a long-standing issue. 
First evidence was reported by the SELEX experiment for 
 $\Xi_{cc}^{++}$ decaying into $ \Lambda_c^+ K^- \pi^+ $ and $pD^+K^-  $  in final states \cite{Mattson:2002vu,Ocherashvili:2004hi}.
The mass measured by SELEX, averaged over the two decay
modes, was found to be $ 3518.7\pm 1.7~\text{MeV}/c^2 $.
However, this has not been confirmed by any other experiments so far.
The FOCUS \cite{Ratti:2003ez}, BaBar \cite{Aubert:2006qw}, LHCb \cite{Aaij:2013voa} and Belle \cite{Kato:2013ynr} experiments
did not find any evidence up to 2017.
In 2017, the doubly charmed baryon $ \Xi^{++}_{cc} $ was
observed by the LHCb collaboration via the decay
channel $ \Xi^{++}_{cc}\rightarrow\Lambda_c^+ K^- \pi^+\pi^−$ \cite{Aaij:2017ueg}, 
and confirmed via measuring another decay channel $\Xi^{++}_{cc}\rightarrow \Xi^+_c \pi^+$ \cite{Aaij:2018gfl}.
The weighted average of its mass for the two decay 
modes was determined to be 
$3621.24\pm 0.65 (\text{stat.})\pm 0.31 (\text{syst.})~\text{MeV}/c^2 $.
Recently,
with a data sample corresponding to an integrated luminosity
 of 9 \ensuremath{\mbox{fb}^{-1}} at the centre-of-mass
energies of 7, 8 and 13 TeV, the LHCb 
Collaboration published the results of a search for the doubly charmed
baryon  $\Xi^+_{cc}$ \cite{Aaij:2019jfq}.
 The upper limit of the ratio of the production
cross-sections between the $ \Xi^+_{cc} $ and $ \Lambda_c^+ $
 baryons times the branching fraction of the $ \Xi^{++}_{cc}\rightarrow\Lambda_c^+ K^- \pi^+$
 decay, was improved by an order of magnitude than the previous search. 
 However, still no significant signal is observed in 
 the mass range from $ 3.4 $  to $3.8~\text{GeV}/c^2  $.
Future LHCb searches with further improved trigger conditions, 
additional $ \Xi^+_{cc} $ decay modes, and larger data
samples should significantly increase the $ \Xi^+_{cc} $ signal sensitivity.

Theoretical studies on the properties and nature of the doubly heavy baryons can play an important role in searching for other members and help us get useful knowledge on the internal structures of the observed resonances.
There have been many theoretical  efforts aimed at understanding the properties of the doubly-heavy baryon states, see e.g. Refs. \cite{Aliev:2012ru,Aliev:2012iv,Azizi:2014jxa,Wang:2017mqp,Wang:2017azm,
Gutsche:2017hux,Li:2017pxa,Xiao:2017udy,Sharma:2017txj,Ma:2017nik,Hu:2017dzi,Shi:2017dto,Yao:2018zze,Yao:2018ifh,Zhao:2018mrg,Wang:2018lhz,
Liu:2018euh,Xing:2018lre,Dhir:2018twm,Berezhnoy:2018bde,Jiang:2018oak,Zhang:2018llc,Gutsche:2018msz,Shi:2019fph,Hu:2019bqj,Brodsky:2011zs,Yan:2018zdt,Cheng:2020wmk,Ozdem:2018uue}. However, most researches are focused on the mass and weak decays of the doubly heavy baryons and the number of studies dedicated to  their strong  decays and the strong couplings of these baryons with other hadrons is very limited.
In this context, we investigate the strong coupling constants among the doubly heavy $\Xi_{cc}$/$\Xi_{bb}$/$\Xi_{bc}$ baryons and $\pi$ mesons. We use the well established non-perturbtive method of light cone QCD sum rule  (LCSR) (for more about this
method see, e.g., \cite{Chernyak:1990ag, Khodjamirian:1997ub,Bagan:1997bp,Ball:1998tj,Ball:2002ps} and references therein) as a powerful theoretical tool to calculate the coupling constants under study. In the calculations, we use the general forms of the interpolating currents for $\Xi_{cc}$, $\Xi_{bc}$  and $\Xi_{bb}$ baryons and DAs of the pions.

The outline of the paper is as follows. In Sec.~\ref{LH}, the light cone sum rules for the
coupling constants of the doubly heavy baryons with  $ \pi $ mesons are obtained.
In Sec.~\ref{NA}, we
present the numerical results and discussions, and Sec.~\ref{SC} is reserved for our conclusions.

\section{THEORETICAL FRAMEWORK}\label{LH}
The starting point is to choose a suitable correlation function (CF) in terms of hadronic currents sandwiched between the QCD vacuum and the on-shell pseudoscalar meson (here the pion). The QCD vacuum interacts via vacuum fluctuations with the initial and final states and leads to non-perturbative contributions to the final results via the quark-quark, quark-gluon and gluon-gluon condensates. The DAs of $\pi$ meson also contain non-perturbative information.

To calculate the physical observables like strong coupling constants, we have to calculate the CF in the large timelike momenta by inserting the full set of hadronic states in the CF and isolating the ground state from the continuum and excited states. On the other hand, in QCD or  theoretical side, we need to study the CF in the large space-like momenta via the operator product expansion (OPE) that separates the perturbative and non-perturbative contributions in terms of different  operators and distribution amplitudes of the particles under consideration.  
Thus, in QCD side, by contracting the hadronic currents in terms of quark fields we  write the CF in terms of the light and heavy quarks propagators as well as wavefunctions of the considered meson in $  x$-space. 
To proceed, we perform the Fourier transformation to transfer the calculations to the momentum space.
To suppress the unwanted contributions coming from the higher states and continuum, we apply the Borel transformation as well as continuum subtraction, supplied by the quark-hadron duality assumption. The two representations of the same CF are then connected to each other via dispersion  integrals. These procedures introduce some auxiliary parameters to the calculations, which are then fixed based on the standard  prescriptions of the method. 
\subsection*{\textit{Correlation Function}}
In order to calculate the strong coupling constants among doubly heavy baryons  and
light pseudoscalar  $ \pi $ meson, we start our discussion by considering the following light-cone correlation function:
 \begin{eqnarray}\label{equ1} 
\Pi(p,q)= i \int d^4x e^{ipx} \left< {\cal P}(q) \vert {\cal T} \left\{
\eta (x) \bar{\eta} (0) \right\} \vert 0 \right>~,
\end{eqnarray}
where $ {\cal P}(q) $ denotes the pseudoscalar mesons of momentum $ q $, $ {\cal T} $ 
is the time ordering operator, and
$ \eta $ represents the interpolating currents of the doubly heavy baryons.
The general expressions of the interpolating currents for the spin–1/2 doubly heavy baryons in their symmetric and antisymmetric forms can be written as
\begin{eqnarray}
 \eta^{S}&=&\frac{1}{\sqrt{2}}\epsilon_{abc}\Bigg\{(Q^{aT}Cq^b)\gamma_{5}Q'^c+
(Q'^{aT}Cq^b)\gamma_{5}Q^c+t (Q^{aT}C\gamma_{5}q^b)Q'^c+t(Q'^{aT}C
\gamma_{5}q^b)Q^c\Bigg\},\nonumber\\
\eta^{A}&=&\frac{1}{\sqrt{6}}\epsilon_{abc}\Bigg\{2(Q^{aT}CQ'^b)\gamma_{5}q^c+
(Q^{aT}Cq^b)\gamma_{5}Q'^c-(Q'^{aT}Cq^b)\gamma_{5}Q^c+2t (Q^{aT}C
\gamma_{5}Q'^b)q^c\nonumber\\
&+& t(Q^{aT}C\gamma_{5}q^b)Q'^c-t(Q'^{aT}C\gamma_{5}q^b)Q^c\Bigg\},
\end{eqnarray}
where $t$ is an arbitrary auxiliary parameter and
the  case,  $t=-1$ corresponds to the Ioffe current.
Here $Q^{(')}$ and $q$ stand for the heavy and
light quarks respectively; $a$, $b$, and $c$ are the color indices, 
$C$ stands for the charge conjugation operator and $T$ denotes the transposition.
For the doubly heavy baryons with two identical heavy quarks, the antisymmetric form of the interpolating  current is zero and we just need to employ the symmetric form, $\eta^{S}$.
In the following, we calculate the
correlation function in Eq.~\ref{equ1} in  two different windows.

\subsection*{\textit{Physical Side} }
To obtain the physical side of  correlation
function, we insert  complete sets of
hadronic states with the same quantum numbers as the interpolating currents and isolate the
ground state. After performing the integration over four-$  x$, we get
\begin{eqnarray}
	\Pi^{\text{Phys.}}(p,q)=\frac{\langle 0\vert \eta\vert B_2(p)\rangle  \langle B_2(p){\cal P}(q)\vert B_1(p+q)\rangle\langle B_1(p+q) \vert \bar{\eta}\vert 0\rangle}{(p^2-m_1^2)[(p+q)^2-m_2^2]} +\cdots~,
\end{eqnarray}
where dots in the above equation stand for the contribution of the  higher states and continuum. 
To proceed we introduce  the matrix elements for spin-1/2 baryons as
\begin{eqnarray}
	\langle 0\vert \eta\vert B_2(p,r)\rangle &=&\lambda_{B_2}u(p,r) ~,\nonumber\\
	\langle B_1(p+q,s) \vert \bar{\eta}\vert 0\rangle &=& \lambda_{B_1}\bar{u}(p+q,s)~,\nonumber\\
	\langle B_2(p,r){\cal P}(q)\vert B_1(p+q,s)\rangle &=& g_{B_1 B_2{\cal P}}
	\bar{u}(p,r)\gamma_5 
	u(p+q,s) ~,
\end{eqnarray}
where $ g_{B_1 B_2{\cal P}} $, representing the strong decay $ B_1\rightarrow B_2 {\cal P}$, is the strong coupling constant among the baryons $ B_1 $ and $B_2  $ as well as the $ {\cal P} $ meson, $\lambda_{B_1}$ and $\lambda_{B_2}$   are the residues of the corresponding baryons and $u(q,s)  $ is Dirac spinor with spin $ s $. 
Putting the above equations all together, and performing summation 
over spins, we get the following representation of the correlator for the phenomenological side:
\begin{eqnarray}\label{CFPhys}
\Pi^{\text{Phys.}}(p,q)=\frac{ g_{B_1 B_2{\cal P}}\lambda_{B_1}\lambda_{B_2}}{(p^2-m_{B_2}^2)[(p+q)^2-m_{B_1}^2]}[\rlap/q \rlap/p\gamma_5 +\cdots~],
\end{eqnarray}
where dots represent other structures come from spin summation as well as the contributions of higher states and continuum. We will use the explicitly presented structure to extract the value of the strong coupling constant, $ g_{B_1 B_2{\cal P}} $.
We apply the double Borel transformation with respect to the variables
$ p^2_1=(p+q)^2 $ and $ p^2_2=p^2 $:
\begin{eqnarray}
{\cal B}_{p_1}(M_1^2){\cal B}_{p_2}(M_2^2)\Pi^{\text{Phys.}}(p,q) &\equiv & \Pi^{\text{Phys.}}(M^2)\nonumber\\
&=& g_{B_1 B_2 {\cal P}} \lambda_{B_1} \lambda_{B_2} e^{-m_{B_1}^2/M_1^2} e^{-m_{B_2}^2/M_2^2}\rlap/q \rlap/p\gamma_5~ + \cdots~,
\end{eqnarray}
 where
$ M^2= M^2_1 M^2_2/(M^2_1+M^2_2) $ and the Borel parameters $ M^2_1 $ and $ M^2_2 $ for the problem under consideration are chosen to be equal as the masses of the initial and final state baryons are the same. Hence $ M^2_1 = M^2_2 = 2M^2$.
\subsection*{\textit{QCD Side}}
In QCD side, the correlation function is calculated in deep Euclidean region with the
help of OPE.
To proceed, we need to determine the correlation function 
using the quark propagators and distribution amplitudes
of the $ \pi $ meson.
The $ \Pi^{QCD}(p,q) $ can be written in the following general form:
\begin{equation}\label{QCD1}
\Pi^{\text{QCD}}(p,q)=\Pi\big((p+q)^2,p^2\big)  \rlap/q \rlap/p\gamma_5+ \cdots~,
\end{equation}
where the $\Pi\big((p+q)^2,p^2\big)$ is an invariant function that should be calculated in terms of QCD degrees of freedom as well as the parameters inside the DAs.

Inserting, for instance, $\eta^{S}$ in the CF and using the Wick theorem to contract all the heavy quark fields we get the following expression in terms of the heavy quark propagators and $\pi$ meson matrix elements:
\begin{eqnarray}\label{QCD2}
\Pi^{\text{QCD}}_{\rho\sigma}(p,q) &=&  \frac{i}{2}\epsilon_{abc}\epsilon_{a'b'c'} \int d^4 x e^{i q.x} \langle\pi(q) \vert \bar{q}^{c^\prime}_{\alpha}(0)q^{c}_{\beta}(x)\vert 0\rangle \Bigg\{ \Bigg[ \Big(\tilde{S}^{aa^{\prime}}_{Q}(x) \Big)_{\alpha\beta}\Big( \gamma_5 S^{bb^{\prime}}_{Q^{\prime}}(x) \gamma_5\Big)_{\rho\sigma}\nonumber \\
&+& \Big(\gamma_5 S_{Q^{\prime}}^{bb^{\prime}}(x)C\Big)_{\rho\alpha}\Big(CS^{aa^{\prime}}_{Q}(x)\gamma_5 \Big)_{\beta\sigma} + t\Big\{\Big(\gamma_5 \tilde{S}_{Q}^{aa^{\prime}}(x)\Big)_{\alpha\beta}\Big(\gamma_5 S^{bb^{\prime}}_{Q^{\prime}}(x) \Big)_{\rho\sigma} \nonumber\\
&+& \Big( \tilde{S}_{Q}^{aa^{\prime}}(x)\gamma_5\Big)_{\alpha\beta}\Big(S^{bb^{\prime}}_{Q^{\prime}}(x)\gamma_5 \Big)_{\rho\sigma} +\Big(\gamma_5 S_{Q^{\prime}}^{bb^{\prime}}(x)C\gamma_5\Big)_{\rho\alpha}\Big(CS^{aa^{\prime}}_{Q}(x)\Big)_{\beta\sigma} \nonumber\\
&-&\Big( S_{Q^{\prime}}^{bb^{\prime}}(x)C\Big)_{\rho\alpha}\Big(\gamma_5CS^{aa^{\prime}}_{Q}(x)\gamma_5\Big)_{\beta\sigma}
 \Big\}+ t^2 \Big\{\Big(\gamma_5 \tilde{S}_{Q}^{aa^{\prime}}(x)\gamma_5\Big)_{\alpha\beta}\Big(S^{bb^{\prime}}_{Q^{\prime}}(x)\Big)_{\rho\sigma} \nonumber\\
  &-& \Big( S_{Q^{\prime}}^{bb^{\prime}}(x)C\gamma_5\Big)_{\rho\alpha}\Big(\gamma_5CS^{aa^{\prime}}_{Q}(x)\Big)_{\beta\sigma} \Big\}\Bigg]+\Bigg(Q \longleftrightarrow Q^{\prime}\Bigg)\Bigg\},
\end{eqnarray}
where $\tilde{S}=C S^T C$, and $\langle \pi(q) \vert \bar{q}^{b^\prime}_{\alpha}(0)q^{b}_{\beta}(x)\vert 0\rangle$ are the matrix elements for the light quark contents of the doubly heavy baryons.
To proceed, we need to know the explicit expression for the heavy quark propagator that is 
\begin{eqnarray}\label{HQP}
S_Q(x) &=& {m_Q^2 \over 4 \pi^2} {K_1(m_Q\sqrt{-x^2}) \over \sqrt{-x^2}} -
i {m_Q^2 \rlap/{x} \over 4 \pi^2 x^2} K_2(m_Q\sqrt{-x^2})\nnb \\& -&
ig_s \int {d^4k \over (2\pi)^4} e^{-ikx} \int_0^1
du \Bigg[ {\rlap/k+m_Q \over 2 (m_Q^2-k^2)^2} G^{\mu\nu} (ux)
\sigma_{\mu\nu} +
{u \over m_Q^2-k^2} x_\mu G^{\mu\nu} \gamma_\nu \Bigg]+...~,\nnb \\
\end{eqnarray}
where $K_1$ and $K_2$ are the modified Bessel functions of the second kind, and $G_{ab}^{\mu \nu }\equiv G_{A}^{\mu \nu }t_{ab}^{A}$ with $A=1,\,2\,\ldots 8$, and  $t^{A}=\lambda ^{A}/2$, where $\lambda ^{A}$ are the Gell-Mann matrices.
The first two terms correspond to perturbative or free part and the rest belong to the interacting parts.

The next step is to use  the heavy quark propagator and the matrix elements $ \langle \pi(q) \vert \bar{q}^{b^\prime}_{\alpha}(0)q^{b}_{\beta}(x)\vert 0\rangle $ in Eq.~\ref{QCD2}.
This leads to different kinds of contributions to the CF. 
 Figs.\ \ref{fig:PertD} and \ref{fig:Npert1} are the Feynman diagrams correspond to the leading and next-to-leading order contributions, respectively which are considered in this work. Only matrix elements corresponding to these diagrams are available.
To calculate the leading order contribution, the heavy quark propagators are replaced by just their perturbative parts. This contribution can be computed using $\pi$ meson two-particle DAs of twist two and higher.  

\begin{figure}[h]
\centering
\includegraphics[width=0.47\textwidth]{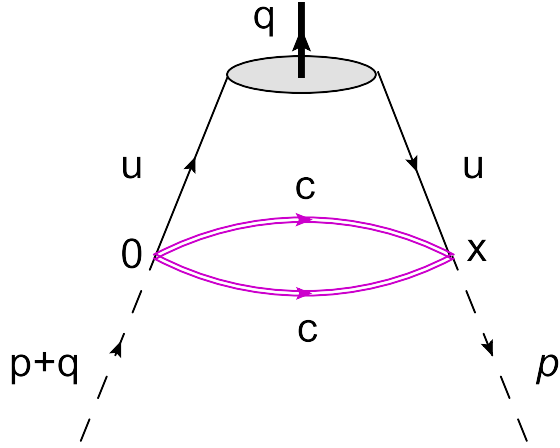}
\caption{The leading order diagram contributing to QCD side.}
\label{fig:PertD}
\end{figure}

\begin{figure}[h]
	\centering
	\includegraphics[width=0.47\textwidth]{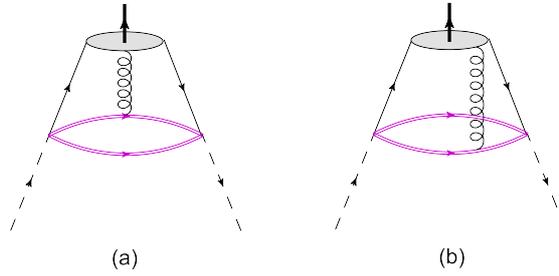}
	\caption{The one-gluon exchange diagrams corresponding to the next-to-leading contributions.}
	\label{fig:Npert1}
\end{figure}

The next-to-leading order contributions can also be calculated by choosing the gluonic parts in Eq.~\ref{HQP} for one of the heavy quark propagators and leaving the other with its perturbative term. They can be expressed in terms of pion three particles DAs. The terms involving more than one gluon field that proportional to four-particle DAs or more are neglected as they are not available 

Now, we concentrate on the strong decay $\Xi_{cc}^{++}\rightarrow \Xi_{cc}^{++}\pi^0$ with the aim of calculating the corresponding strong coupling constant $ g_{\Xi_{cc}^{++} \Xi_{cc}^{++}\pi^0} $. The other channels have  similar procedures. 
To proceed, we replace the heavy quark propagators in \ref{QCD2}  by their explicit expression and perform the summation over the color indices by applying 
the replacement
\begin{equation}\label{sumcolor1}
\overline{u}_{\alpha }^{a}(x)u_{\beta }^{a^{\prime }}(0)\rightarrow \frac{1}{%
	3}\delta _{aa^{\prime }}\overline{u}_{\alpha }(x)u_{\beta }(0).
\end{equation}
Now, using the expression 
\begin{equation}
\overline{u}_{\alpha }(u)u_{\beta }(0)\equiv \frac{1}{4}\Gamma_{\beta \alpha }^{J}\overline{u}(x)\Gamma^{J}u(0),  \label{eq:Expan}
\end{equation}
one can relate the CF to the DAs of the pion with different twists. Here the summation over $J$ runs as 
\begin{equation}
\Gamma ^{J}=\mathbf{1,\ }\gamma _{5},\ \gamma _{\mu },\ i\gamma _{5}\gamma
_{\mu },\ \sigma _{\mu \nu }/\sqrt{2}.
\end{equation}%
Following the similar way one can calculate the contributions involving the gluon field.

As a result, the CF is found in terms of the QCD parameters as well as the matrix elements
\begin{eqnarray}
&&\langle \pi^0 (q)|\overline{u}(x)\Gamma ^{J}u(0)|0\rangle ,  \notag \\
&&\langle \pi^0 (q)|\overline{u}(x)\Gamma ^{J}G_{\mu \nu }(vx)u(0)|0\rangle ,
\label{eq:MatElem}
\end{eqnarray}%
whose expressions in terms of the wave functions of the pion with different twists are given in the Appendix.

Inserting the expression of the above-mentioned matrix elements in term of wave functions of different twists we get the CF in 
$ x $ space. This is followed by the Fourier and Borel transformations as well as continuum subtraction.
To proceed we need to perform the Fourier transformation of the following kind:

\begin{eqnarray}\label{STR1}
T_{[~~,\alpha,\alpha\beta]}(p,q)&=& i \int d^4 x \int_{0}^{1} dv  \int {\cal D}\alpha e^{ip.x} \big(x^2 \big)^n  [e^{i (\alpha_{\bar q} + v \alpha _g) q.x} \mathcal{G}(\alpha_{i}) , e^{iq.x} f(u)] \nonumber\\
  &\times& [1 , x_{\alpha} , x_{\alpha}x_{\beta}]  K_{\mu}(m_Q\sqrt{-x^2})  K_{\nu}(m_Q\sqrt{-x^2}),
\end{eqnarray}
where the expressions in the brackets denote different possibilities arise in the calculations, the blank subscript in the left hand side indicates no indices regarding no $x_\alpha$ in the configuration,  $\mathcal{G}(\alpha_{i})$ and $f(u)$ represent wave functions coming from the three and two-particle matrix elements
 and $ n $ is a positive integer. The measure
\begin{equation*}
\int \mathcal{D}\alpha =\int_{0}^{1}d\alpha _{q}\int_{0}^{1}d\alpha _{\bar{q}%
}\int_{0}^{1}d\alpha _{g}\delta (1-\alpha _{q}-\alpha _{\bar{q}}-\alpha
_{g}),
\end{equation*}
is used in the calculations. To start the Fourier  transformation, we use
\begin{equation}\label{trick1}
(x^2)^n = (-1)^n \frac{d^n}{d \beta^n}\big(e^{- \beta x^2}\big)\arrowvert_{\beta = 0}.
\end{equation}
for positive integer $ n $ and
\begin{eqnarray}\label{trick2}
x_{\alpha} e^{i P.x} = (-i) \frac{d}{d P^{\alpha}} e^{i P.x}.
\end{eqnarray}
We also use the following representation of  the Bessel functions $K_{\nu }$ (see also Ref. \cite{Azizi:2018duk}):
\begin{equation}\label{CosineRep}
K_\nu(m_Q\sqrt{-x^2})=\frac{\Gamma(\nu+ 1/2)~2^\nu}{\sqrt{\pi}m_Q^\nu}\int_0^\infty dt~\cos(m_Qt)\frac{(\sqrt{-x^2})^\nu}{(t^2-x^2)^{\nu+1/2}}.
\end{equation}

As an example let us consider the following generic form: 
\begin{eqnarray}\label{Z1}
{\cal Z}_{\alpha\beta}(p,q) &=& i \int d^4 x \int_{0}^{1} dv  \int {\cal D}\alpha e^{i[p+ (\alpha_{\bar q} + v \alpha _g)q].x} \mathcal{G}(\alpha_{i}) \big(x^2 \big)^n  \nonumber\\
&\times& x_\alpha x_\beta  K_{\mu}(m_Q\sqrt{-x^2})  K_{\nu}(m_Q\sqrt{-x^2}).
\end{eqnarray}
We substitute Eqs. \ref{trick1}, \ref{trick2} and \ref{CosineRep} into \ref{Z1}. Then to perform the $x$-integration we go to the Euclidean space by Wick rotation and  get
\begin{eqnarray}\label{STR2}
{\cal Z}_{\alpha\beta}(p,q) &=& \frac{i  \pi^2 2^{\mu+\nu-2}}{m_{Q_1}^{2\mu} m_{Q_2}^{2\nu}}\int  \mathcal{D}\alpha  \int_{0}^{1} dv \int_{0}^{1} dy_1 \int_{0}^{1} dy_2 \frac{\partial }{\partial P_{\alpha}} \frac{\partial }{\partial P_{\beta}} \frac{\partial^n }{\partial \beta^n}\nonumber\\
&\times& \dfrac{y_{1}^{\mu-1} y_{2}^{\nu-1}}{(y_1+y_2+\beta)^2}  e^{-\frac{1}{4} \big(\frac{m_{Q_1}^{2}}{y_1} + \frac{m_{Q_2}^{2}}{y_2} - \frac{P^2}{y_1+y_2+\beta}\big)},
\end{eqnarray}
where $P=p+q(v \alpha_{g} +\alpha_{q})$ .
Changing variables from $y_1$ and $y_2$ to $\rho$ and $z$ as 
\begin{equation} \label{ChangeVar}
~~~~~~~~~~~\rho=y_1+y_2,~~~~~~~~z=\frac{y_1}{y_1+y_2},
\end{equation}
and taking derivative with respect to $P_{\alpha}$ and $P_{\beta}$, we get

\begin{eqnarray}\label{STR3}
{\cal Z}_{\alpha\beta}(p,q) &=& \frac{i  \pi^2 2^{\mu+\nu-4}}{m_{Q_1}^{2\mu} m_{Q_2}^{2\nu}}\int  \mathcal{D}\alpha  \int_{0}^{1} dv \int_{0}^{\infty} d\rho \int_{0}^{1} dz \frac{\partial^n }{\partial \beta^n} z^{\mu-1}\bar{z}^{\nu-1} \frac{\rho^{\nu+\mu-1}}{(\rho+\beta)^4} e^{-\frac{1}{4} \big(\frac{m_{Q_1}^{2} \bar{z} + m_{Q_2}^{2}z }{z \bar{z} \rho} + \frac{P^2}{\beta +\rho}\big)} \nonumber\\
&\times& \Big[ p_\alpha p_\beta + (v \alpha_{g} +\alpha_{q})(p_\alpha q_\beta +q_\alpha p_\beta ) + (v \alpha_{g} +\alpha_{q})^2 q_\alpha q_\beta + 2 (\rho + \beta)g_{\alpha\beta} \Big].
\end{eqnarray}
Now we perform the double Borel transformation using
\begin{equation} \label{Borel1}
{\cal B}_{p_1}(M_{1}^{2}){\cal B}_{p_2}(M_{2}^{2})e^{b (p + u q)^2}=M^2 \delta(b+\frac{1}{M^2})\delta(u_0 - u) e^{\frac{-q^2}{M_{1}^{2}+M_{2}^{2}}},
\end{equation}
where  $u_0 = M_{1}^{2}/(M_{1}^{2}+M_{2}^{2})$. After integrating over $\rho$ we have
\begin{eqnarray}\label{STR4}
{\cal Z}_{\alpha\beta}(M^2) &=& \frac{i  \pi^2 2^{4-\mu-\nu} e^{\frac{-q^2}{M_1^2+M_2^2}}}{M^2 m_{Q_1}^{2\mu} m_{Q_2}^{2\nu}}\int  \mathcal{D}\alpha  \int_{0}^{1} dv \int_{0}^{1} dz \frac{\partial^n }{\partial \beta^n} e^{-\frac{m_{Q_1}^2 \bar{z} + m_{Q_2}^2 z}{z \bar{z}(M^2 - 4\beta)}} z^{\mu-1}\bar{z}^{\nu-1} (M^2 - 4\beta)^{\mu+\nu-1} \nonumber\\
&\times & \delta[u_0 - (\alpha_{q} + v \alpha_{g})]  \Big[ p_\alpha p_\beta + (v \alpha_{g} +\alpha_{q})(p_\alpha q_\beta +q_\alpha p_\beta ) + (v \alpha_{g} +\alpha_{q})^2 q_\alpha q_\beta \nonumber \\ 
&&+ \frac{M^2}{2}g_{\alpha\beta} \Big].
\end{eqnarray}

 The next step is to perform the continuum subtraction in order to more suppress the contribution of the higher states and continuum. The subtraction procedures for different systems are described in Ref.\ \cite{Azizi:2018duk} in details. When the masses of the initial and final baryonic states are equal, as we stated previously, we set $M_{1}^{2}=M_{2}^{2}=2M^{2}$. In this case, the double spectral density  is
concentrated  near the diagonal $s_1=s_2$ and reduces to a single representation  $ s $ (see also Ref. \cite{Agaev:2016srl} and references therein) and for the continuum subtraction more simple expressions are  derived, which are not sensitive to the shape of the
duality region. For the case, $M_{1}^{2}=M_{2}^{2}=2M^{2}$ and $u_{0}=1/2$,  the subtraction procedure is explained in Ref. \cite{Agaev:2016srl} in details, which we use in the calculations.

 
By  calculation of all the Fourier integrals and applying the Borel transformation and continuum subtraction, for QCD side of the calculations in Borel Scheme, we get,
\begin{eqnarray}
\Pi^{QCD}(M^2,s_0,t)=\big[\Pi^{(0)}(M^2,s_0,t) + \Pi^{(GG)}(M^2,s_0,t)\big] \rlap/q \rlap/p \gamma_{5}+ \cdots,
\end{eqnarray}
where the functions  $  \Pi^{(0)}(M^2,s_0,t) $ and   $  \Pi^{(GG)}(M^2,s_0,t) $ are obtained as
\begin{eqnarray}\label{fbspert}
\Pi^{(0)}(M^2,s_0,t) &=& \dfrac{ e^{-\frac{m_{\pi}^2}{4M^2}}}{96 \pi^2\sqrt{2}} \int_{0}^{1}dz \frac{1}{z}e^{\frac{-m_c^2}{M^2 z \bar{z}}}\Bigg\{ 3f_{\pi} m_{\pi}^2 m_c^3 (t^2-1)  {\mathbb A}(u_0)  \nonumber \\
&+& 2e^{-\frac{4m_c^2}{M^2}}  \int_{4m_c^2}^{s_0} dse^{-\frac{s}{M^2}}z\bigg[ 3f_{\pi} m_{\pi}^2 m_c (t^2-1)\bar{z} {\mathbb A}(u_0) \nonumber \\
&-& 6f_{\pi}m_c(s-4m_c^2)(t^2-1)\bar{z} \varphi_{\pi}(u_0) \nonumber \\ 
&-& \mu_{\pi}(\tilde \mu_{\pi}^2 -1)\big[-4m_c^2t+3(s-4m_c^2)(t-1)^2 z\bar{z} \big]\varphi_{\sigma}(u_0) \bigg] \nonumber \\
&+& 6 e^{-\frac{4m_c^2}{M^2}}(t^2-1)\int_{4m_c^2}^{s_0} ds e^{-\frac{s}{M^2}} \int_{0}^{1} dv \int \mathcal{D}\alpha  \\
&\times&  \Bigg[f_{\pi}m_{\pi}^2 m_c\delta[u_0-(\alpha_{q}+v\alpha_{g})]\bigg( -2z(v-1/2) {\cal A}_\parallel (\alpha_i) + 2\bar{z}{\cal V}_\perp(\alpha_i) \nonumber \\
&+& (2z-3){\cal V}_\parallel(\alpha_i)  \bigg) + 2 \mu_\pi \delta^{\prime}[u_0-(\alpha_{q}+v\alpha_{g})] (s-4m_c^2) z \bar{z} (v-1/2) {\cal T}(\alpha_i)\Bigg]\Bigg\}, \nonumber
\end{eqnarray}
and
\begin{eqnarray}
	\Pi^{(GG)}(M^2,s_0,t) &=& \frac{\langle g_s^2G^2 \rangle e^{-\frac{m_{\pi}^2}{4 M^2}}}{6912 \sqrt{2} \pi^2 m_c M^6 }\int_{0}^{1}dz \frac{1}{z^2 \bar{z}^4}e^{-\frac{m_c^2}{M^2 z \bar{z}}} \nonumber \\
	&\times&\Bigg\{\bar{z}^2 \Bigg[ -3f_{\pi} m_{\pi}^2(1-t^2) \Big( 6 M^6 z^2 \bar{z}^4+6 M^4 m_c^2 z \bar{z}^3+3 M^2 m_c^4 \bar{z}^2-2 m_c^6 \Big) {\mathbb A}(u_0) \nonumber \\
	&+& 4M^2 m_c \bar{z} \Bigg( -3 f_\pi M^2 m_c (1-t^2)z \big[ 2m_c^2 + M^2 \bar{z}(5z-3) \big]\varphi_{\pi}(u_0) \nonumber\\
	&+& (\tilde \mu_{\pi}^2 -1)\mu_\pi \Big[ 2 m_c^4 \left(t^2+1\right)+ [(1+t^2)(1-4z)-6t](M^2 m_c^2 z + M^4 z^2 \bar{z}) \Big]\varphi_{\sigma}(u_0) \Bigg)  \nonumber \\
	&+&  72 f_\pi M^8 (1-t^2)z^2 \bar{z}^4 \Big( 1-e^{-\frac{(s_0-4m_c^2)}{M^2}} \Big)\varphi_{\pi}(u_0) \Bigg]\nonumber \\ 
	&+& 6M^2(1-t) \int_{0}^{1}dv \int {\cal D}\alpha \Bigg[ f_\pi m_\pi^2 (1 + 
	t) \bar{z}^3\delta[u_0-(\alpha_{q}+v\alpha_{g})]\Bigg( 2 m_c^4 {\cal V}_\perp(\alpha_i) \nonumber \\
	&+& \Big[-2 m_c^4 + M^2 m_c^2(1 + 2 z) z  + 
	M^4  (1 + 2 z)z^2 \bar{z}\Big] {\cal V}_\parallel(\alpha_i) \nonumber \\ &-&(2v-1) \Big[m_c^4 + M^2 m_c^2 (1 + 2 z)z  + 
	M^4 (1 + 2 z)z^2 \bar{z}\Big] {\cal A}_\parallel (\alpha_i) \Bigg) \nonumber \\ 
	&+&  \mu_{\pi} (1 - t) (1 + 2 v) z \bar{z}^3 \delta^{\prime}[u_0-(\alpha_{q}+v\alpha_{g})] (M^2 m_c^3 + M^4 m_c z \bar{z})  {\cal T}(\alpha_i) \Bigg]\Bigg\}.
	\end{eqnarray}

The sum rule for the coupling constant under study is found by matching the coefficients
of the structure $ \rlap/q \rlap/p \gamma_{5} $ from both the physical and QCD
sides.  As a result, we get:
\begin{eqnarray}\label{SR}
g_{B_1 B_2  {\cal P}}(M^2,s_0,t)=\frac{1}{\lambda_{\Xi_{cc}}\lambda_{\Xi_{cc}}} e^{\frac{m_{B_1}^2}{2M^2}}e^{\frac{m_{B_2}^2}{2M^2}}\big[\Pi^{(0)}(M^2,s_0,t) + \Pi^{(GG)}(M^2,s_0,t)\big].
\end{eqnarray}
As is seen,  the sum rules for coupling constants  contain the residues of doubly heavy baryons, which are borrowed from  Ref.\cite{Aliev:2012ru}. 
Similarly,  we obtain the sum rules   for other  strong coupling constants under consideration.

\section{NUMERICAL ANALYSIS}\label{NA}

In this section, we numerically analyze the sum rules for the strong coupling constants of
the $ \pi $ mesons with $ \Xi_{cc} $, $\Xi_{bc}$ and $ \Xi_{bb}$ baryons and discuss the 
results. The sum rules for the couplings $g_{\Xi_{cc} \Xi_{cc} \pi}  $ , $g_{\Xi_{bc} \Xi_{bc} \pi}$ and $g_{\Xi_{bb} \Xi_{bb} \pi}  $ contain some input parameters like, quark masses the mass and decay constant of the $ \pi $ meson and the masses and residues of doubly heavy baryons.
 They were extracted from experimental data or calculated from  nonperturbative methods.
The values of some of these parameters  together with quark masses are given in Tables \ref{tab:pi}. As we previously mentioned the values of the residues of baryons are used from Ref.\cite{Aliev:2012ru}.
\begin{table}[t]
\renewcommand{\arraystretch}{1.3}
\addtolength{\arraycolsep}{1pt}
$$
\begin{tabular}{|c|c|c|c|}
\hline \hline
       \mbox{Parameters}         &  \mbox{Values}  
\\
\hline\hline
  $ m_{c} $    &  $ 1.27\pm0.02~\mbox{GeV} $      \\
    $ m_b $    & $  4.18^{+0.03}_{-0.02}~\mbox{GeV} $      \\
 $ m_{\pi^{0} } $    &  $ 134.98~\mbox{MeV} $      \\
$ m_{\pi^{\pm}  }  $  &  $ 139.57~\mbox{MeV} $      \\
 $ \Xi_{cc}  $    &  $ 3621.2\pm0.7~\mbox{MeV} $    \\
 $ \Xi_{bc} $    &  $ 6.72\pm0.20~\mbox{GeV}  $ \cite{Aliev:2012ru}    \\
  $ \Xi_{bb} $    &  $ 9.96\pm0.90~\mbox{GeV}  $ \cite{Aliev:2012ru}    \\
$  f_\pi $    & $  130.2\pm1.2~\mbox{MeV}  $     \\
 \hline \hline
\end{tabular}
$$
\caption{Some input values used in the calculations. They  are mainly taken from \cite{Tanabashi:2018oca}, except the ones that the references are given next to the numbers.}  \label{tab:pi} 
 \renewcommand{\arraystretch}{1}
\addtolength{\arraycolsep}{-1.0pt}
\end{table} 
Another set of important  input parameters are the
$ \pi $ meson wavefunctions of different twists, entering the DAs. These wavefunctions  are given as  \cite{Ball:1998je,Ball:2004ye}:
\begin{eqnarray}
\phi_{\pi}(u) &=& 6 u \bar u \Big( 1 + a_1^{\pi} C_1(2 u -1) + a_2^{\pi} C_2^{3 \over 2}(2 u - 1) \Big), 
\nonumber \\
{\cal T}(\alpha_i) &=& 360 \eta_3 \alpha_{\bar q} \alpha_q \alpha_g^2 \Big( 1 + w_3 \frac12 (7 \alpha_g-3) \Big),
\nonumber \\
\phi_P(u) &=& 1 + \Big( 30 \eta_3 - \frac{5}{2} \frac{1}{\mu_{\pi}^2}\Big) C_2^{1 \over 2}(2 u - 1) 
\nonumber \\ 
&+&	\Big( -3 \eta_3 w_3  - \frac{27}{20} \frac{1}{\mu_{\pi}^2} - \frac{81}{10} \frac{1}{\mu_{\pi}^2} a_2^{\pi} \Big) C_4^{1\over2}(2u-1),
\nonumber \\
\phi_\sigma(u) &=& 6 u \bar u \Big[ 1 + \Big(5 \eta_3 - \frac12 \eta_3 w_3 - \frac{7}{20}  \mu_{\pi}^2 - \frac{3}{5} \mu_{\pi}^2 a_2^{\pi} \Big)
C_2^{3\over2}(2u-1) \Big],
\nonumber \\
{\cal V}_\parallel(\alpha_i) &=& 120 \alpha_q \alpha_{\bar q} \alpha_g \Big( v_{00} + v_{10} (3 \alpha_g -1) \Big),
\nonumber \\
{\cal A}_\parallel(\alpha_i) &=& 120 \alpha_q \alpha_{\bar q} \alpha_g \Big( 0 + a_{10} (\alpha_q - \alpha_{\bar q})\Big),
\nonumber \\
{\cal V}_\perp (\alpha_i) &=& - 30 \alpha_g^2\Big[ h_{00}(1-\alpha_g) + h_{01} (\alpha_g(1-\alpha_g)- 6 \alpha_q \alpha_{\bar q}) +
	h_{10}(\alpha_g(1-\alpha_g) - \frac32 (\alpha_{\bar q}^2+ \alpha_q^2)) \Big],
\nonumber \\
{\cal A}_\perp (\alpha_i) &=& 30 \alpha_g^2(\alpha_{\bar q} - \alpha_q) \Big[ h_{00} + h_{01} \alpha_g + \frac12 h_{10}(5 \alpha_g-3) \Big],
\nonumber \\
B(u)&=& g_{\pi}(u) - \phi_{\pi}(u),
\nonumber \\
g_{\pi}(u) &=& g_0 C_0^{\frac12}(2 u - 1) + g_2 C_2^{\frac12}(2 u - 1) + g_4 C_4^{\frac12}(2 u - 1),
\nonumber \\
{\mathbb A}(u) &=& 6 u \bar u \left[\frac{16}{15} + \frac{24}{35} a_2^{\pi}+ 20 \eta_3 + \frac{20}{9} \eta_4 +
	\Big( - \frac{1}{15}+ \frac{1}{16}- \frac{7}{27}\eta_3 w_3 - \frac{10}{27} \eta_4 \right) C_2^{3 \over 2}(2 u - 1) 
	 \nonumber \\ 
	&+&\Big( - \frac{11}{210}a_2^{\pi} - \frac{4}{135} \eta_3w_3 \Big)C_4^{3 \over 2}(2 u - 1)\Big]
\nonumber \\
&+& \Big( -\frac{18}{5} a_2^{\pi} + 21 \eta_4 w_4 \Big)\Big[ 2 u^3 (10 - 15 u + 6 u^2) \ln u 
  \nonumber \\
&+& 2 \bar u^3 (10 - 15 \bar u + 6 \bar u ^2) \ln\bar u + u \bar u (2 + 13 u \bar u) \Big],
\end{eqnarray}
where $C_n^k(x)$ are the Gegenbauer polynomials and 
\begin{eqnarray}
h_{00}&=& v_{00} = - \frac13\eta_4,
\nonumber \\
a_{10} &=& \frac{21}{8} \eta_4 w_4 - \frac{9}{20} a_2^{\pi},
\nonumber \\
v_{10} &=& \frac{21}{8} \eta_4 w_4,
\nonumber \\
h_{01} &=& \frac74  \eta_4 w_4  - \frac{3}{20} a_2^{\pi},
\nonumber \\
h_{10} &=& \frac74 \eta_4 w_4 + \frac{3}{20} a_2^{\pi},
\nonumber \\
g_0 &=& 1,
\nonumber \\
g_2 &=& 1 + \frac{18}{7} a_2^{\pi} + 60 \eta_3  + \frac{20}{3} \eta_4,
\nonumber \\
g_4 &=&  - \frac{9}{28} a_2^{\pi} - 6 \eta_3 w_3.
\label{param0}
\end{eqnarray}
The constants inside the wavefunctions are calculated at
 the renormalization scale of $\mu=1 ~\mbox{GeV}^{2}$ and they are given as
$a_{1}^{\pi} = 0$, $a_{2}^{\pi} = 0.44$,
$\eta_{3} =0.015$, $\eta_{4}=10$, $w_{3} = -3$ and 
$ w_{4}= 0.2$ \cite{Ball:1998je,Ball:2004ye}.

Finally, the sum rules for the coupling constants contain 
three auxiliary parameters: Borel mass parameter 
$ M^2 $,  continuum threshold $ s_0 $ and the general parameter $  t$ entered the
general spin–$ 1/2 $ currents. We should find working regions
of these parameters, at which the results of coupling constants 
have relatively small variations with respect to the changes of these parameters.
To restrict these parameters, we employ the standard prescriptions of the method such
as the pole dominance, convergence  of the OPE and  mild variations of the physical quantities
with respect to the auxiliary parameters.  The upper limit of $ M^2 $ is determined from the pole dominance condition, i.e.,
\begin{equation}
\frac{\Pi^{QCD}(M^2,s_0,t)}{\Pi^{QCD}(M^2,\infty,t)}>\frac{1}{2}.
\end{equation}
The lower limit of $ M^2 $ is fixed by the  condition of OPE convergence:  in our case,   $  \Pi^{(0)}(M^2,s_0,t) >    \Pi^{(GG)}(M^2,s_0,t) $.
The continuum threshold $s_0$ is not totally  arbitrary and it  depends on the mass of the first excited state in the same channel. One has to choose the range of $s_0$ such that it does not contain the energy for producing the first excited state. Unfortunately, there is no experimental information on the masses of the first excited states in the case of doubly heavy baryons. Based on our analyses and considering the experimental information on the single heavy baryons, we consider the interval  $m_{QQ}+E_1\leq\sqrt{s_0}\leq  m_{QQ}+E_2$ for $\sqrt{s_0}  $, where a energy from $ E_1 $ to $ E_2 $ is needed to excite the baryons, and  impose  that the Borel curves are  flat and the requirements of the  pole dominance and the OPE convergence  are satisfied. With these criteria, we choose the $s_0$ to lie  in the interval $(m_{\Xi_{QQ}}+0.3)^2\leq s_0\leq (m_{\Xi_{QQ}}+0.7)^2 (\text{GeV}^2)$.

As a result of the above requirements, we  obtain the working region of
the Borel parameter for the $ \Xi_{cc}$ channel as,
\begin{equation}
3~\mbox{\rm GeV}^2 \leq M^2 \leq 6 ~\mbox{\rm GeV}^2.
\end{equation}
The continuum threshold  for this channel is
obtained as,
\begin{equation}
16 ~\mbox{\rm GeV}^2 \leq s_0 \leq 18 ~\mbox{\rm GeV}^2.
\end{equation}
For $\Xi_{bc}$ baryon, we get
\begin{equation}
8~\mbox{\rm GeV}^2 \leq M^2 \leq 12~\mbox{\rm GeV}^2~, ~~~
49 ~\mbox{\rm GeV}^2 \leq s_0 \leq 55 ~\mbox{\rm GeV}^2~.
\end{equation}
Finally, for $\Xi_{bb}$ baryon, these  parameters lie in the intervals:
\begin{equation}
18~\mbox{\rm GeV}^2 \leq M^2 \leq 24~\mbox{\rm GeV}^2, ~~~
106 ~\mbox{\rm GeV}^2 \leq s_0 \leq 114 ~\mbox{\rm GeV}^2~.
\end{equation}
The working window for the parameter $t$ is  obtained by the consideration of the minimum variations of the results with respect to this parameter.  By imposing this condition together with the conditions of the pole dominance and convergence of the mass sum rules the working window for $ t $ is obtained in Ref. \cite{Aliev:2012ru} as $|t| \leq 2$, which we also use in our analyses.
 As examples, we display the dependence  of the strong coupling constant
$g_{\Xi_{cc}^{++} \Xi_{cc}^{++} \pi^0}  $,  which is obtained from the sum rule for the strong coupling form factor at $ q^2=m^2_{\pi} $, with respect to $ M^2 $ and $ s_0 $ in Figs.~\ref{fig:msq2} and \ref{fig:s0} at $ t=-2 $.
\begin{figure}[h!]
\centering
\includegraphics[width=0.50\textwidth]{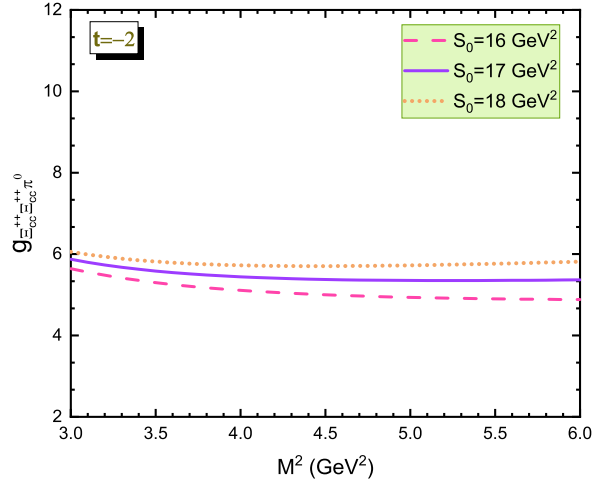}
\caption{The strong coupling $g_{\Xi_{cc}^{++} \Xi_{cc}^{++} \pi^0}  $ 
as a function of the Borel
parameter $ M^2 $ at $ t=-2 $ for different values of $ s_0 $.}
\label{fig:msq2}
\end{figure}

\begin{figure}[h!]
\centering
\includegraphics[width=0.50\textwidth]{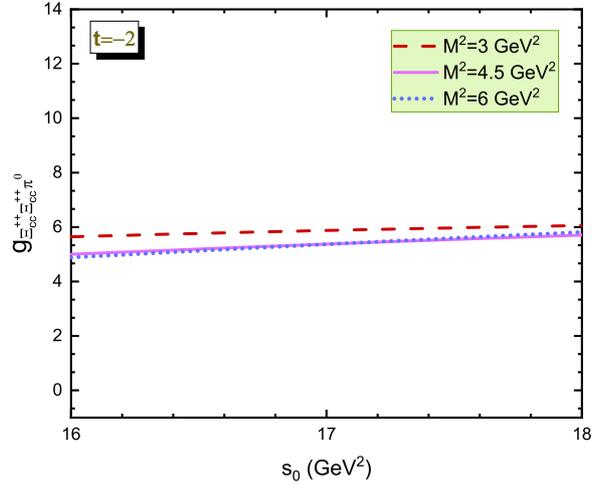}
\caption{The strong coupling $g_{\Xi_{cc}^{++} \Xi_{cc}^{++} \pi^0}  $ 
as a function of  $ s_0$ at $t=-2 $ for different values of $M^2 $.}
\label{fig:s0}
\end{figure}
From these figures we see mild variations of $g_{\Xi_{cc} \Xi_{cc} \pi^0}  $ 
with respect to the $M^2$ and $s_0$, which appear as the main uncertainty
in the numerical values of the strong coupling constants. We extract the numerical values of the  strong couplings  $g_{\Xi_{cc}\Xi_{cc} \pi^0 } $, $g_{\Xi_{bc} \Xi_{bc} \pi^0}  $  and  $g_{\Xi_{bb} \Xi_{bb} \pi^0}  $  as  displayed in table ~\ref{tab:g}. The presented errors are due to the changes with respect to the auxiliary 
parameters in their working regions as well as those which propagate from other input parameters as 
well as $ \pi $ meson DAs. The values of coupling constants to the charged mesons $ \pi^+ $ and $ \pi^ -$,  which are exactly the same from the isospin symmetry, are found by the multiplications of the strong coupling constants in $ \pi^0 $ channel by $ \sqrt{2} $. This coefficient is the only difference in the couplings to the quark contents of the  $ \pi^\pm $ and  $ \pi^0 $ mesons when the isospin symmetry is used. As it is also clear from table \ref{tab:g},  the values of the strong coupling constants in double-$ b $ channel are roughly four times greater than those of the other channels. The big difference between the strong couplings to pseudoscalar mesons  in $ b $ and $ c$ channels is evident in the case of single heavy baryons, as well \cite{Aliev:2010yx}.  As it can be seen from this reference,  the   difference factor  in the case of single heavy baryons is two-three times. 
\begin{table}[h!]
	\renewcommand{\arraystretch}{1.3}
	\addtolength{\arraycolsep}{1pt}
	$$
	\begin{array}{|c|c|c|c|c|c|}
	\hline \hline
	\mbox{Channel}          &  \mbox{strong coupling constant}  
	\\
	\hline\hline
	\Xi_{cc}\rightarrow \Xi_{cc} \pi^0   & 5.52^{\:+0.64}_{\:-0.53}   \\
	\Xi_{bc}\rightarrow  \Xi_{bc} \pi^0    & 4.75^{\:+0.42}_{\:-0.50}   \\
	\Xi_{bb}\rightarrow  \Xi_{bb} \pi^0    & 21.60^{\:+1.77}_{\:-2.09}   \\
	\hline \hline
	\end{array}
	$$
	\caption{
		The numerical values for the strong coupling constants extracted from the analyses.} \label{tab:g} 
	\renewcommand{\arraystretch}{1}
	\addtolength{\arraycolsep}{-1.0pt}
\end{table}
 
 \newpage

\section{SUMMARY AND CONCLUSIONS}\label{SC}
The doubly charmed $\Xi_{cc}^{++} (ccu)$ baryon is the only listed doubly heavy baryon in PDG discovered in the experiment so far. The LHCb collaboration gets closer to observing other member $\Xi_{cc}^{+} (ccd)$, as well. Therefore, the investigation of the doubly charmed/bottom baryons from many aspects is of great importance that may help us in the course of search for new members of the doubly heavy baryons predicted by the quark model. The strong coupling constants among the hadronic multiplets are fundamental objects that can help us to explore the nature and structure of the participating particles as well as the properties of QCD as the theory of strong interaction.

We calculated the  strong coupling constants $g_{\Xi_{ccq} \Xi_{ccq} \pi^{0,\pm}}  $, $g_{\Xi_{bcq} \Xi_{bcq} \pi^{0,\pm}} $ and $g_{\Xi_{bbq} \Xi_{bbq} \pi^{0,\pm}}  $, with $ q $ being either $ u $  or $ d $ quark, in the framework of the light cone QCD sum rule and using the general form of the interpolating currents for the doubly heavy baryons and the $\pi  $ meson's DAs. Based on the standard prescriptions of the method, we fixed the auxiliary parameters entering the calculations. 
We extracted the values of the strong coupling constants at different channels. Our results may be checked via different theoretical models and approaches. The obtained results may help us in constructing the strong interaction potential among the doubly heavy baryons and the pseudoscalar mesons. Our results may also help experimental groups in analyses of the obtained related data in  hadron colliders. 


\appendix
\section*{Appendix: The pion distribution amplitudes} \label{APA}
In this appendix, we present explicit expressions for  the DAs  of the $\pi  $ meson. For more information see Refs.~ \cite{Ball:1998je,Ball:2004ye}.
\begin{eqnarray}
\langle {\pi}(p)| \bar q(x) \gamma_\mu \gamma_5 q(0)| 0 \rangle &=& -i f_{\pi} p_\mu  \int_0^1 du  e^{i \bar u p x} 
	\left( \varphi_{\pi}(u) + \frac{1}{16} m_{\pi}^2 x^2 {\mathbb A}(u) \right)
\nonumber \\
	&-& \frac{i}{2} f_{\pi} m_{\pi}^2 \frac{x_\mu}{px} \int_0^1 du e^{i \bar u px} {\mathbb B}(u), 
\nonumber \\
\langle {\pi}(p)| \bar q(x) i \gamma_5 q(0)| 0 \rangle &=& \mu_{\pi} \int_0^1 du e^{i \bar u px} \varphi_P(u),
\nonumber \\
\langle {\pi}(p)| \bar q(x) \sigma_{\alpha \beta} \gamma_5 q(0)| 0 \rangle &=& 
\frac{i}{6} \mu_{\pi} \left( 1 - \tilde \mu_{\pi}^2 \right) \left( p_\alpha x_\beta - p_\beta x_\alpha\right)
	\int_0^1 du e^{i \bar u px} \varphi_\sigma(u),
\nonumber \\
\langle {\pi}(p)| \bar q(x) \sigma_{\mu \nu} \gamma_5 g_s G_{\alpha \beta}(v x) q(0)| 0 \rangle &=&
	i \mu_{\pi} \left[
		p_\alpha p_\mu \left( g_{\nu \beta} - \frac{1}{px}(p_\nu x_\beta + p_\beta x_\nu) \right) 
\right. \nonumber \\
	&-&	p_\alpha p_\nu \left( g_{\mu \beta} - \frac{1}{px}(p_\mu x_\beta + p_\beta x_\mu) \right) 
\nonumber \\
	&-&	p_\beta p_\mu \left( g_{\nu \alpha} - \frac{1}{px}(p_\nu x_\alpha + p_\alpha x_\nu) \right)
\nonumber \\ 
	&+&	p_\beta p_\nu \left. \left( g_{\mu \alpha} - \frac{1}{px}(p_\mu x_\alpha + p_\alpha x_\mu) \right)
		\right]
\nonumber \\
	&\times& \int {\cal D} \alpha e^{i (\alpha_{\bar q} + v \alpha_g) px} {\cal T}(\alpha_i),
\nonumber \\
\langle {\pi}(p)| \bar q(x) \gamma_\mu \gamma_5 g_s G_{\alpha \beta} (v x) q(0)| 0 \rangle &=& 
	p_\mu (p_\alpha x_\beta - p_\beta x_\alpha) \frac{1}{px} f_{\pi} m_{\pi}^2 
		\int {\cal D}\alpha e^{i (\alpha_{\bar q} + v \alpha_g) px} {\cal A}_\parallel (\alpha_i)
\nonumber \\
	&+& \left[
		p_\beta \left( g_{\mu \alpha} - \frac{1}{px}(p_\mu x_\alpha + p_\alpha x_\mu) \right) \right.
\nonumber \\
	&-& 	p_\alpha \left. \left(g_{\mu \beta}  - \frac{1}{px}(p_\mu x_\beta + p_\beta x_\mu) \right) \right]
	f_{\pi} m_{\pi}^2
\nonumber \\
	&\times& \int {\cal D}\alpha e^{i (\alpha_{\bar q} + v \alpha _g) p x} {\cal A}_\perp(\alpha_i),
\nonumber \\
\langle {\pi}(p)| \bar q(x) \gamma_\mu i g_s G_{\alpha \beta} (v x) q(0)| 0 \rangle &=& 
	p_\mu (p_\alpha x_\beta - p_\beta x_\alpha) \frac{1}{px} f_{\pi} m_{\pi}^2 
		\int {\cal D}\alpha e^{i (\alpha_{\bar q} + v \alpha_g) px} {\cal V}_\parallel (\alpha_i)
\nonumber \\
	&+& \left[
		p_\beta \left( g_{\mu \alpha} - \frac{1}{px}(p_\mu x_\alpha + p_\alpha x_\mu) \right) \right.
\nonumber \\
	&-& 	p_\alpha \left. \left(g_{\mu \beta}  - \frac{1}{px}(p_\mu x_\beta + p_\beta x_\mu) \right) \right]
	f_{\pi} m_{\pi}^2
\nonumber \\
	&\times& \int {\cal D}\alpha e^{i (\alpha_{\bar q} + v \alpha _g) p x} {\cal V}_\perp(\alpha_i),
\end{eqnarray}
where 
\begin{eqnarray}
\label{nolabel}
\mu_{ \pi} = f_{\pi} {m_{\pi}^2\over m_{u} + m_{d}}~,~~~~~ 
\widetilde{\mu}_{\pi} = {m_{u} + m_{d} \over m_{\pi}}.
\end{eqnarray}
 Here,
 $\varphi_{\pi}(u),$ $\Bbb{A}(u),$ $\Bbb{B}(u),$ 
$\varphi_P(u),$ $\varphi_\sigma(u),$ 
${\cal T}(\alpha_i),$ ${\cal A}_\perp(\alpha_i),$ ${\cal A}_\parallel(\alpha_i),$ 
${\cal V}_\perp(\alpha_i)$ and ${\cal V}_\parallel(\alpha_i)$
are wave functions of definite twists.

\end{document}